# Towards Direct-Gap Silicon Phases by the Inverse Band Structure Design Approach

H. J. Xiang[1,2*], Bing Huang[2], Erjun Kan[3], Su-Huai Wei[2], X. G. Gong[1]

[1] Key Laboratory of Computational Physical Sciences (Ministry of Education), State Key Laboratory of Surface Physics, and Department of Physics, Fudan University, Shanghai 200433, P. R. China

[2] National Renewable Energy Laboratory, Golden, Colorado 80401, USA

[3] Department of Applied Physics, Nanjing University of Science and Technology, Nanjing, Jiangsu 210094, People's Republic of China

e-mail: hxiang@fudan.edu.cn

**Abstract**

Diamond silicon (Si) is the leading material in current solar cell market. However, diamond Si is an indirect band gap semiconductor with a large energy difference (2.4 eV) between the direct gap and the indirect gap, which makes it an inefficient absorber of light. In this work, we develop a novel inverse-band structure design approach based on the particle swarming optimization algorithm to predict the metastable Si phases with better optical properties than diamond Si. Using our new method, we predict a cubic $Si_{20}$ phase with quasi-direct gaps of 1.55 eV, which is a promising candidate for making thin-film solar cells.

Due to the high stability, high abundance, and the existence of an excellent compatible oxide ($SiO_2$), Si is the leading material of microelectronic devices. Currently, the majority of solar cells fabricated to date have also been based on diamond Si in monocrystalline or large-grained polycrystalline form [1]. There are mainly two reasons for this: First, Si is the second most abundant element in the earth's crust; Second, the Si based photovoltaics (PV) industry could benefit from the successful Si based microelectronics industry. It is well known that Si is an indirect band gap semiconductor with a large energy difference between the direct gap (3.5 eV) and the indirect gap (1.1 eV). Due to the indirect nature of diamond Si, the Si solar cell absorber layer should be thick enough to absorb all the low energy photons that has to be assisted by phonon. To reduce the Si material usage and thus the cost, it is desirable to discover new Si phases which are more direct than diamond Si.

The optical properties of several high-pressure metastable phase of silicon were studied theoretically by Cohen and co-workers: The GW calculation showed that Si-XII (R8 structure) exhibits an indirect band gap of approximately 0.24 eV at ambient pressure [2]; Silicon in the lonesdaleite phase has an indirect band gap of 0.95 eV [2]. It can be seen that known metastable phases of Si are not good material for solar cell absorber. Another way to improve the efficiency of the Si-based solar cell is through alloying Si with other elements. For example, it was shown [3] that homogenous $Si_3AlP$ alloy has a larger fundamental band gap (1.41 eV) and a smaller direct band gap (about 2.00 eV) than diamond Si, suggesting that $Si_3AlP$ is a better material for solar cell absorber due to the increased open-circuit voltage and short-circuit current. However, the direct band gap of $Si_3AlP$ is still far from the optimal value (1.5 eV) for solar cell applications.

In this Letter, we aim at predicting new metastable phases of silicon with better optical properties. To achieve this, we first develop an inverse-band structure design approach based on the particle swarming optimization (PSO) algorithm. With the newly developed approach, we predict a new cubic Si phase with a quasi-direct dipole transition allowed gap of 1.55 eV. We suggest that this new Si phases could be used in thin-film solar cells.

The inverse band structure design [4] is a theoretical method that addresses the problem of finding the atomic configuration of a system having a target electronic-structure property. Usually, the inverse band structure design approach was used to design the configurations of the ordered alloy (with the fixed lattice structure) which have the target electronic-structure property. For instance, the AlGaAs alloys for maximum bandgap and minimum bandgap were predicted [5] by the inverse band structure design approach based on the genetic algorithm. Here the problem we are encountering is different: Which Si metastable phase has the direct bandgap and the allowed band edge optical transition. The shape of the unit cells and lattice structures are not known in prior. To address this issue, we develop a new inverse band structure design approach based on the PSO algorithm as implemented in the CALYPSO code for the structure prediction [6]. PSO is designed to solve problems related to multidimensional optimization, which is inspired by the social behavior of birds flocking or fish schooling. Recently, the PSO algorithm was adopted to predict new crystal structures [6-8] with the lowest enthalpy at given external conditions (e.g., pressure). To the best of our knowledge, the PSO algorithm has never been applied to the inverse band structure design problem. The reason why we choose the PSO algorithm but not the genetic algorithm is that there is no crossover operation in the PSO algorithm. In the case of predicting lowest energy structures, the cut-and-splice crossover operation [9-11] in the genetic algorithm is efficient. However, the cut-and-splice crossover operation becomes inefficient in the current case because the electronic structures such as the band gap are global properties of the system, which depend not just on the local structural motifs.

In our PSO algorithm for the inverse band structure design, we first generate $N_p$ random structures with randomly selected space groups. Subsequently, local optimization including the atomic coordinates and lattice parameters is performed for each of the initial structures. For each of the relaxed structures, we then compute the electronic structure using a dense k-mesh, and the transition matrix elements between the valence band maximum (VBM) and conduction band minimal (CBM) at each k-point are evaluated. From the electronic structure, we can determine the indirect gap ($E_g^{id}$) and direct gap ($E_g^d$) of the system. If the indirect gap is the same as the direct gap, then the system is a direct gap system. The fitness of a structure is defined as: $f = E_g^d + w \times (E_g^{id} - E_g^d) + t$, where $w$ chosen to be 8 is a weight parameter, and $t$ is $-10$ if the transition at the direct gap is forbidden. The definition of the fitness function is to favor the structure with an optically active larger direct band gap. We then update the pbest and gbest according to the fitness. Some of the structures of the next generations are generated by the PSO operation: $x_{i,j}^{t+1} = x_{i,j}^t + v_{i,j}^{t+1}$. The new velocity of each particle is calculated on the basis of its previous location $x_{i,j}^t$, previous velocity $v_{i,j}^t$, current location $\text{pbest}_{i,j}^t$ with an achieved best fitness, and the population global location $\text{gbest}_{i,j}^t$ with the best fitness value for the entire population. It should be noted that in the implementation of CALYPSO [6], pbest for the *k*-th particle in the population is defined as the locally optimized structure of the *k*-th structure when predicting the lowest energy structures. Here, we use the traditional definition for pbest in the original PSO algorithm for the inverse band structure design: For the *k*-th particle in the population, pbest is defined as the structure with the largest fitness among all the structures that the k-th particle has visited. The other structures are generated randomly, which is critical to increase the structure diversity. Our implementation of the PSO algorithm for the inverse band structure design is based on the PSO implementation in CALYPSO code [6]. The local density approximation (LDA) functional [12] is used in the structural relaxation and the calculation of the optical spectra in the PSO simulation. For the final accurate calculations of the optical spectra of the $Si_{20}$ phase, we adopt the HSE06 functional. The density functional calculations are performed by using the VASP code [13]. To check the efficiency of our method, we try to search a $TiO_2$ structure which has a predefined band gap (i.e., the band gap of rutile $TiO_2$). Our three test simulations show that we can successfully find the rutile structure in two steps on average.

We mainly consider the structures with different number of Si atoms in the unit cell: $N_a = 2 \times m$, where $m$ ranges from 1 to 10. Here we only consider the structures with even number of Si atoms because it is most unlikely that a Si structure with the odd number of atoms has good optical properties. We have performed some tests on $Si_3$ and $Si_5$ to confirm that no direct gap system is found. In our PSO simulations, we usually set the population size to 24. The number of generations is fixed to 30. For each $N_a$, we repeat the calculations for several times.

From our simulation, we find a new structure with 20 atoms (see [14] for detailed structural parameters) in the unit cell which has good optical properties. The $Si_{20}$ structure [see Fig. 1(a), denoted as $Si_{20}$-T] has the cubic T symmetry with space group No. 198 ($P2_13$). It has three kinds of inequivalent Si atoms: one 12b Si position, and two 4a Si positions. Every Si atom is four-fold coordinated: Each 4a Si atom bonds with three 12b Si atoms and another 4a Si atom, while each 12b Si atom has two neighboring 12b Si atoms and two inequivalent 4a Si atoms. However, the Si tetrahedrons are distorted. In particular, the 12a Si atoms form equilateral triangles with each other.

We use the HSE06 functional [15] to calculate the electronic structures of the $Si_{20}$ phase because the HSE06 functional was shown to predict much better electronic properties than the local or semi-local density approximations. $Si_{20}$-T has a quasi-direct band gap [see Fig. 2(a)] near (0.17,0.17,0.17). The direct band gap of $Si_{20}$-T is 1.55 eV, which is much larger than that of diamond Si. The overall VBM locates near (0,0.25,0), but is just 0.06 eV higher than the VBM at (0.17,0.17,0.17). Some states at special k-points (e.g., $\Gamma$) are two-fold or three-fold degenerate because of the cubic point group T. Our G0W0 [16] calculations based on the HSE06 wavefunctions and eigenvalues predict that the fundamental band gap for $Si_{20}$-T is 1.61 eV. This indicates that HSE06 indeed predicts accurate band gaps for Si phases, as also found for the diamond Si case [17].

The imaginary part of dielectric function for $Si_{20}$-T from the HSE06 calculations is shown in Fig. 3. For comparison, we also show the imaginary part of dielectric function for diamond Si. The optical absorption in $Si_{20}$-T also starts at the direct gap transition energy, i.e., 1.5 eV. Therefore, the direct gap transition in $Si_{20}$-T is dipole allowed. It is well-know that a semiconductor with a direct gap of 1.5 eV is most suitable for the use as solar absorption material. $Si_{20}$-T has a large fundamental band gap than diamond Si, and it has a smaller direct band gap than diamond Si. The increase of the fundamental band gap of $Si_{20}$-T than Si is beneficial in increasing the open-circuit voltage and the decrease of the direct optical band gap is beneficial in increasing the absorption, thus, the photocurrent of the solar cell. Therefore, we propose that $Si_{20}$-T could be a better solar cell absorber than diamond Si. In particular, $Si_{20}$-T could be used to make thin-film solar cells because of the allowed low energy direct transitions.

We now try to understand why the $Si_{20}$ phase has good optical properties. For $Si_{20}$-T, the point group for the direct gap k-point (0.17,0.17,0.17) is $C_3$ with the three-fold axis along the [111] direction. The $C_3$ point group has three irreducible representations (A, E and $E^*$), all of which are one-dimensional. We find that the VBM state belong to the E representation, while the CBM state belong to the A representation. Thus, the dipole transition between the VBM and CBM state along the direction perpendicular to the [111] direction is allowed. As shown in Fig. 4(a) and 4(b), the VBM and CBM states distribute mostly around the triangles composed by 12b Si atoms. Therefore, the presence of Si triangles seems to be important to the optical transitions in $Si_{20}$-T.

It is well-known that diamond Si is the most stable phase for Si. The new $Si_{20}$ phase is less stable than diamond Si by about 0.3 eV/Si due to the distortion of Si tetrahedrons. To examine the dynamic stability of the predicted $Si_{20}$ phase, we compute its phonon dispersion by the finite difference method. As can be seen from Fig. 5(a), there are no imaginary phonon modes in the whole Brillouin zone, indicating that the $Si_{20}$ phase is dynamically stable. The highest frequency of the optical modes is around 550 $cm^{-1}$, which are a little higher than that of diamond Si (508 $cm^{-1}$). It is also necessary to examine whether the $Si_{20}$ phase is thermally stable for the room-temperature solar cell applications. In order to explore this aspect, a large supercell with 160 atoms is built and first-principles molecular dynamic simulations are performed with a Nose-Hoover thermostat at 350 K. Figure 5(b) shows the fluctuations of the temperature and total energy as a function of simulation time. After 13.5 ps, we find no structure destruction of the $Si_{20}$ structure, except for some thermal fluctuations. This shows that the $Si_{20}$ phase is thermal stable up to at least 350 K.

In summary, we have developed a new method for the inverse band structure design based on the PSO algorithm. Our inverse band structure design approach is able to predict new materials with desirable properties without fixing the lattice and structural type. By combining the new method with first principles calculations, we predict a new metastable Si phase with good optical properties: $Si_{20}$-T has a quasi-direct gap of 1.55 eV. We propose that $Si_{20}$-T could be promising solar energy absorber. Experimental synthesis of the new $Si_{20}$ phase is called for to verify our predictions. Our new method for the inverse band structure design can be generally applied to design materials with other desirable physical properties.

**References**

1. Bjørn Petter Jelle, Christer Breivik, and Hilde Drolsum Røkenes, Sol. Energy Mater. Sol. Cells **100**, 69 (2012).

2. B. D. Malone, J. D. Sau, and M. L. Cohen, Phys. Rev. B **78**, 035210 (2008); Phys. Rev. B **78**, 161202 (2008).

3. Ji-Hui Yang, Yingteng Zhai, Hengrui Liu, H. J. Xiang, X. G. Gong, and Su-Huai Wei, J. Am. Chem. Soc. **134**, 12653 (2012).

4. A. Franceschetti and A. Zunger, Nature **402**, 60 (1999); M. d'Avezac, J.-W. Luo, T. Chanier, and A. Zunger, Phys. Rev. Lett. **108**, 027401 (2012).

5. K. Kim, P. A. Graf, and W. B. Jones, J. Comp. Phys. **208**, 735 (2005).

6. Y. Wang, J. Lv, L. Zhu, and Y. Ma, Phys. Rev. B **82**, 094116 (2010); Y. Wang, J. Lv, L. Zhu, and Y. Ma, Comput. Phys. Commun. **183**, 2063–2070 (2012).

7. J. Lv, Y. Wang, L. Zhu, and Y. Ma, Phys. Rev. Lett. **106**, 015503 (2011); Hui Wang, John S. Tse, Kaori Tanaka, Toshiaki Iitaka, and Yanming Ma, Proc. Natl. Acad. Sci. USA **109**, 6463 (2012); Yanchao Wang, Hanyu Liu, Jian Lv, Li Zhu, Hui Wang and Yanming Ma, Nature Commun. **2**, 563 (2011).

8. X. Luo, J. Yang, H. Liu, X. Wu, Y. Wang, Y. Ma, S.-H. Wei, X. G. Gong, and H. J. Xiang, J. Am. Chem. Soc. **133**, 16285 (2011).

9. D. M. Deaven and K. M. Ho, Phys. Rev. Lett. **75**, 288 (1995).

10. A. Oganov and C. Glass, J. Chem. Phys. **124**, 244704 (2006).

11. H. J. Xiang, S.-H. Wei, and X. G. Gong, J. Am. Chem. Soc. **132**, 7355 (2010).

12. J.P. Perdew and A. Zunger, Phys. Rev. B **23**, 5048 (1981); D.M. Ceperley and B.J. Alder, Phys. Rev. Lett. **45**, 566 (1980).

13. G. Kresse and J. Furthmüller, Comput. Mater. Sci. **6**, 15 (1996); Phys. Rev. B **54**, 11169 (1996).

14. See Supplemental Material at http://link.aps.org/supplemental for structural parameters of the new $Si_{20}$ phase.

15. J. Heyd, G. E. Scuseria, and M. Ernzerhof, J. Chem. Phys. **118**, 8207 (2003); J. Chem. Phys. **124**, 219906 (2006).

16. M. S. Hybertsen and S. G. Louie, Phys. Rev. B **34**, 5390 (1986). M. Shishkin and G. Kresse, Phys. Rev. B **74**, 035101 (2006).

17. J. Heyd, J. E. Peralta, G. E. Scuseria, and R. L. Martin, J. Chem. Phys. **123**, 174101 (2005).

**Acknowledgements**

Work at Fudan was partially supported by NSFC, FANEDD, Research Program of Shanghai Municipality and MOE, the Special Funds for Major State Basic Research. Work at NREL is supported by the U.S. Department of Energy under Contract No. DE-AC36-08GO28308. H. X. thanks Dr. Liping Yu for useful discussion.

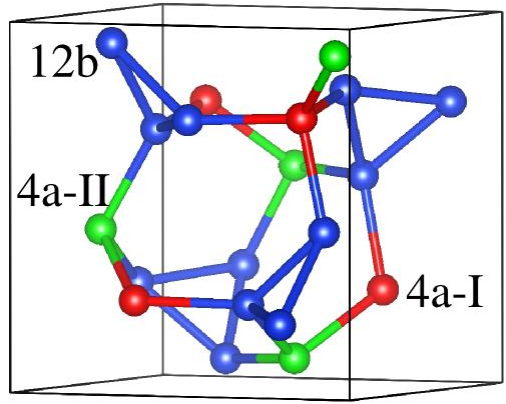

Figure 1. The geometric structure of the new $Si_{20}$ phase ($Si_{20}$-T). The three different Si positions (4a-I, 4a-II, and 12b) are indicated by different colors.

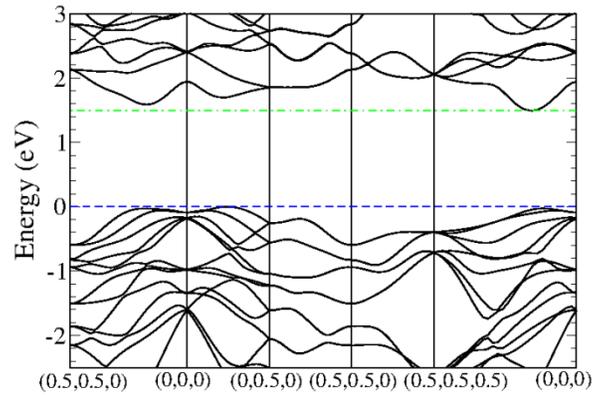

Figure 2. HSE06 band structure for $Si_{20}$-T. The overall VBM and CBM are denoted by dashed line and dot-dashed line, respectively. The k-points below the horizontal axis refer to the fractional coordinates in terms of the reciprocal lattice of the primitive cell.

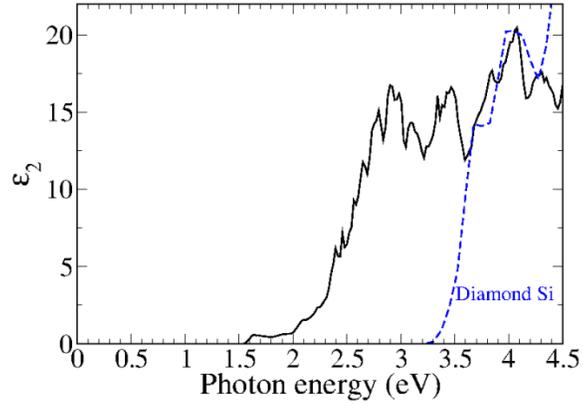

Figure 3. Imaginary part of dielectric functions from the HSE06 calculations for $Si_{20}$-T. For comparison, the imaginary part of dielectric function of diamond Si from the HSE06 calculation is also shown.

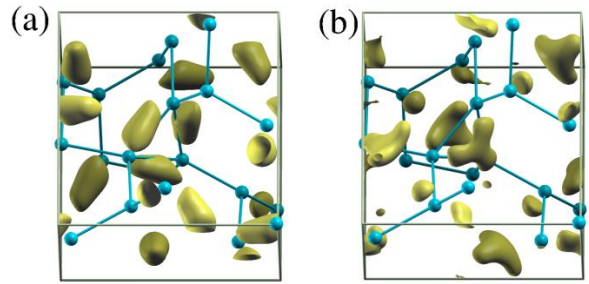

Figure 4. (a) and (b) show the partial charge density of the VBM and CBM states at (0.17,0.17,0.17) for $Si_{20}$-T, respectively. It can be seen that the states distribute mostly around the Si triangles.

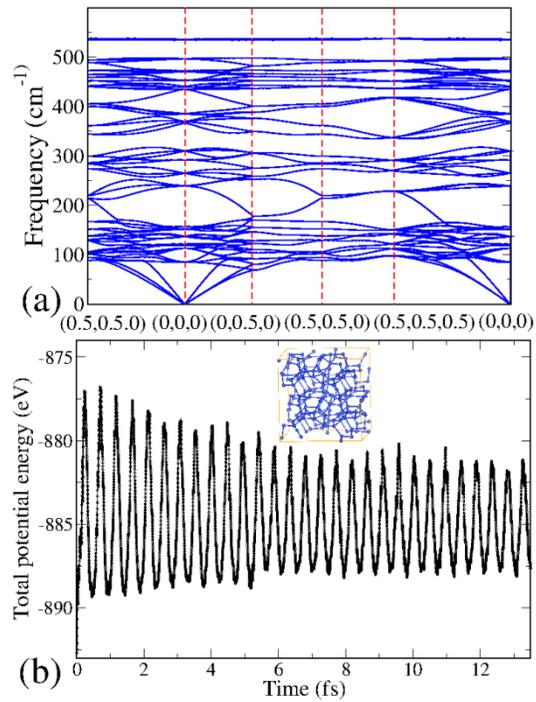

Figure 5. (a) Phonon dispersion for $Si_{20}$-T from the LDA calculations. The q-points are the same as the k-points used in Fig. 2. (b) The fluctuations of potential energy of the $Si_{20}$ supercell as a function of the molecular dynamic simulation step at 350 K. A snapshot of the simulated system is also shown in the inset.